%% file: paper.tex
\documentclass[conference]{IEEEtran}
\IEEEoverridecommandlockouts
\usepackage[
all=normal
,leading=tight
,leadingfraction=0.98
]{savetrees}
\usepackage{cite}
\usepackage{amsmath,amssymb,amsfonts}
\usepackage{algorithm}
\usepackage{amsmath}
\usepackage{amssymb}
\usepackage{amsfonts}
\usepackage{algpseudocode}
\usepackage[hidelinks]{hyperref}
\usepackage{graphicx}
\usepackage{textcomp}
\usepackage{xcolor}
\usepackage{booktabs}
\usepackage{orcidlink}
\usepackage{glossaries}
\usepackage{tcolorbox}
\usepackage{subfig}
\usepackage{lipsum}
\usepackage{caption}
\usepackage{enumitem}
\usepackage{mathtools}
\usepackage{tikz}

\usepackage[detect-all=true,per-mode=symbol,per-symbol =/]{siunitx}
\def\BibTeX{{\rm B\kern-.05em{\sc i\kern-.025em b}\kern-.08em
    T\kern-.1667em\lower.7ex\hbox{E}\kern-.125emX}}

\begin{document}

\bstctlcite{IEEEexample:BSTcontrol}

\input{glossary.tex}

\title{FlatAttention: Dataflow and Fabric Collectives Co-Optimization for Efficient Multi-Head Attention on Tile-Based Many-PE Accelerators}

\author{
    \IEEEauthorblockN{
        Chi Zhang\IEEEauthorrefmark{1},
        Luca Colagrande\IEEEauthorrefmark{1}, 
        Renzo Andri\IEEEauthorrefmark{3},
        Thomas Benz\IEEEauthorrefmark{1},
        Gamze Islamoglu\IEEEauthorrefmark{1}, 
        Alessandro Nadalini\IEEEauthorrefmark{2}, \\
        Francesco Conti\IEEEauthorrefmark{2}, 
        Yawei Li\IEEEauthorrefmark{1}, 
        Luca Benini\IEEEauthorrefmark{1}\IEEEauthorrefmark{2}
    }
    \IEEEauthorblockA{
        \IEEEauthorrefmark{1}Integrated Systems Laboratory (IIS), ETH Zurich, Zurich, Switzerland\\
        \IEEEauthorrefmark{2}Department of Electrical, Electronic, and Information Engineering (DEI), University of Bologna, Bologna, Italy\\
        \IEEEauthorrefmark{3}Computing Systems Lab, Huawei Zurich Research Center, Zurich, Switzerland\\
        \{chizhang, colluca, tbenz, gislamoglu, yawli, lbenini\}@iis.ee.ethz.ch, \\
        \{alessandro.nadalini3, f.conti\}@unibo.it, renzo.andri@huawei.com
    }
}


\newcommand{\ARCH}{Tile-Based Many-PE }
\newcommand{\arch}{tile-based many-PE }
\newcommand{\FlashBW}{80}
\newcommand{\FlatBW}{1.7}
\newcommand{\FlatCollBW}{8.1}
\newcommand{\OptFlatBW}{22.1}
\newcommand{\OptFlatBWSU}{3.8}
\newcommand{\OptFlatHBMTraficSave}{16}
\newcommand{\OptFlatPerfSU}{4.1}
\newcommand{\MaxUti}{89.3}
\newcommand{\MaxEnergySave}{x}
\newcommand{\DieSave}{1.8}
\newcommand{\SUtoNV}{1.3}
\newcommand{\GEMMSUtoNV}{1.2}
\newcommand{\BWreducetoNV}{40}
\newcommand{\redmulekGE}{4536}
\newcommand{\spatzkGE}{3258}
\newcommand{\snitchkGE}{25}
\newcommand{\iDMAkGE}{297}
\newcommand{\routerkGE}{196}

\definecolor{color_ls}{RGB}{151, 180, 104}
\definecolor{color_mc}{RGB}{242, 149,  69}
\definecolor{color_rd}{RGB}{248,  46, 190}
\definecolor{color_cp}{RGB}{218, 118, 113}

\maketitle

\begin{abstract}
Multi-Head Attention (MHA) is a critical computational kernel in transformer-based AI models.
Emerging scalable tile-based accelerator architectures integrate increasing numbers of tightly-packed processing elements (PEs) with tensor units. MHA dataflow mapping is crucial for achieving high utilization of the available units.
We propose FlatAttention, a new dataflow for MHA on tile-based many-PE accelerators, minimizing costly main memory (HBM) accesses by leveraging collective
primitives integrated into the on-chip network fabric. FlatAttention achieves up to \MaxUti\% utilization, and \OptFlatPerfSU\texttimes\ performance speedup over FlashAttention-3 dataflow on tile-based accelerators whilst reducing HBM traffic by \OptFlatHBMTraficSave\texttimes.
Through algorithm-architecture co-exploration, we identify an optimal configuration for a large scaled-out tile-based accelerator featuring a 32×32 tile mesh with 1024 TFLOPS @ FP16 peak performance, comparable to the state-of-the-art Nvidia H100 GPU.
FlatAttention in this configuration achieves up to \SUtoNV× higher utilization over FlashAttention-3 on the H100 GPU.  
Meanwhile, this tile-based accelerator configuration requires \BWreducetoNV\% less HBM bandwidth compared to the H100 GPU, enabling a \DieSave× reduction in die size, estimated on the same technology node.

\end{abstract}


\section{Introduction}
\label{sec:into}

Transformer-based \gls{ai} models, such as GPT-4, LLaMA, and DeepSeek-V3 are dominating \glspl{llm}. 
Among the computational kernels in transformer-based models, \gls{mha} exhibits quadratic complexity over sequence length\cite{keles2023computational}, making it a critical factor in performance, especially for long sequences.
Previous LLM studies \cite{yuan2024llm, ivanov2021data} have shown that most operations in \gls{mha} are bottlenecked by memory accesses.

The ``attention bottleneck'' has driven extensive research focused on optimizing \gls{mha} dataflows on the dominant AI hardware platform, namely Nvidia GPUs.
One of the most widely adopted solutions is FlashAttention~\cite{dao2022}, which efficiently fuses \gls{mha} microkernels.
Over two generations of improvements, FlashAttention-2~\cite{dao2023} introduced algorithmic optimizations, whereas FlashAttention-3~\cite{shah2024} additionally leverages asynchronous execution for improved performance.
However, still no more than 75\%\footnote{FlashAttention-3 baseline. Numbers use arXiv v1 (11 Jul 2024) [FA3-arXiv], the newest version when experiments were run; a later NeurIPS 2024 release reports $\sim$10 \% higher throughput [FA3-NeurIPS].} utilization was achieved on the H100 GPUs~\cite{shah2024}.

Moreover, Nvidia’s \gls{soa} H100 GPU comes with significant cost and power requirements, featuring an 814 mm\textsuperscript{2} die on TSMC's 5nm process node, coupled with six \gls{hbm} stacks accounting for over 50\% of the total cost, and a \gls{tdp} of 700 W.
Given the suboptimal utilization of the \gls{mha} layer and the high cost and power requirements, many competitors are working on hardware and software to improve system cost and energy efficiency with highly optimized accelerators designed for strong \gls{llm} performance.
The goal is to offer competitive performance while boosting the efficiency of the system by minimizing energy-hungry HBM accesses.

As recent application trends, linked to the ``reasoning'' use of LLMs, emphasize inference as a value-added workload, a scalable design pattern for inference accelerators targeting scaled-up transformer-based \gls{ai} models is emerging \cite{prabhakar2024b, lie2024, vasiljevic2024, lee202516}: these accelerators are constructed as large, full-reticle, or even multi-die integrated systems \cite{lie2024} structured as meshes of compute tiles containing extremely dense, large matrix units, coupled with vector and scalar engines to accelerate all key kernels in \glspl{llm}, and local, explicitly managed memories for main memory data buffering and latency hiding.
Several \glspl{hbm} are typically employed as main memory, positioned at die boundaries and supported by multiple memory controllers.

Such a tile-based many-\gls{pe} accelerator architectural template favors silicon efficiency and scalability, featuring a dense compute tile placement and a software-controlled partitioned memory hierarchy.
However, mapping LLM inference workloads onto these scalable tile-based accelerators presents a significant challenge.
The inter-tile and tile-to-HBM dataflow must be carefully designed to achieve high utilization of the tiles' matrix engines and minimize energy-hungry off-chip access. In this work, we show that leveraging collective communication primitives in the \gls{noc}, such as reduction operations and multicast, can be highly beneficial for both of these objectives.

Existing works investigating mapping and architecture co-exploration for LLM workloads either fail to fully consider key optimizations, such as operator fusion and overlap within the \gls{mha} layer, falling short of FlashAttention's highly optimized \gls{soa} dataflow \cite{thuning2024}, or do not explore the use of inter-tile collective primitives in \glspl{noc} \cite{kao2023, nayak2024fusemax}, or both \cite{cai2024}.
Although some AI accelerator vendors claim to achieve high \gls{llm} inference efficiency through optimized dataflows \cite{prabhakar2024b, lie2024}, the details of their dataflow implementations—particularly for the \gls{mha} layer—and on-chip fabric architectures remain confidential. Additionally, accelerator architecture design must be co-explored alongside dataflow optimizations to determine optimal design parameters and features.

This work takes the \gls{mha} in the prefill stage of LLM inference as a case study, which aligns with FlashAttention\cite{dao2022} and could be generalized to training. We propose a new dataflow for scalable tile-based accelerators and present a co-design approach for algorithm-architecture co-exploration. The contributions of this paper are:
\begin{itemize}
    \item A modeling and simulation framework that enables estimating the performance of a large set of tile-based accelerators and the co-design of network primitives.
    \item An efficient workload allocation and scheduling strategy called FlatAttention. FlatAttention leverages collective primitives on the on-chip network fabric to achieve up to \MaxUti\% utilization for \gls{mha} layer on \arch accelerators, and  \OptFlatPerfSU\texttimes\ performance speedup over FlashAttention-3 dataflow on the same tile-based accelerator, whilst reducing HBM traffic by \OptFlatHBMTraficSave\texttimes.
    \item Co-exploration of accelerator architecture and FlatAttention parameters, highlighting key trends and trade-offs to guide the selection of optimal FlatAttention parameters.
    \item An optimal tile-based accelerator configuration that matches the peak performance of the current \gls{soa} Nvidia H100 GPU.
    FlatAttention in our configuration achieves up to \SUtoNV× higher utilization over FlashAttention-3 on H100.
    Meanwhile, this tile-based accelerator configuration requires \BWreducetoNV\% less HBM bandwidth than the H100 while achieving a \DieSave× reduction in die size, estimated on the same technology node.
\end{itemize}

\section{Reference \ARCH Architecture}
\label{sec:arch}

\begin{figure}[t!]
  \centering
  \includegraphics[width=1\linewidth, trim={6cm 5.7cm 8.5cm 4.5cm}, clip]{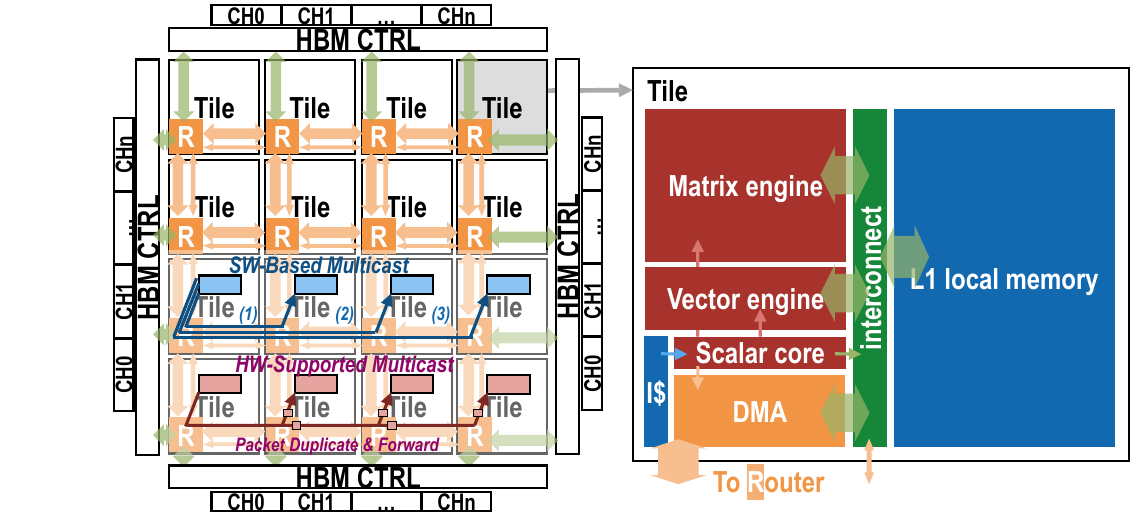}
  \vspace{-3mm}
  \caption{Tile-Based Many-PE Architecture Template}
  \vspace{-6mm}
  \label{fig_arch}
\end{figure}

This section introduces the architecture of the scalable tile-based many-\gls{pe} design pattern, prominently featured in \gls{soa} commercial \gls{ml} accelerators, e.g., Tesla's Dojo system \cite{talpes2022} and Tenstorrent's Blackhole chip \cite{vasiljevic2024}.

As illustrated in Fig. \ref{fig_arch}, the fundamental building block of the many-\gls{pe} architecture is the ``tile''. Each tile comprises \glspl{pe}, local memory (L1), a \gls{dma} engine, and local interconnects. There are three main types of \glspl{pe}: scalar cores, vector engines, and matrix engines. Scalar cores mainly handle dataflow control tasks, whereas heavy computational tasks are offloaded to the vector and matrix engines based on the computation type. All \glspl{pe} within a tile can directly access the local L1 memory via the local interconnect. The \gls{dma} engine in each tile is responsible for data movement in and out of the local L1 memory. 
The tile-based many-\gls{pe} system uses an on-chip 2D-mesh \gls{noc} to connect tiles. Off-chip memory, such as \gls{hbm}, is located at the boundary of the mesh \gls{noc}, interfaced through the respective memory controllers.

Collective communication operations~\cite{vandegeijn2011}, such as multicast and reduction, are involved in L1 data exchange among tiles.
Traditional software-based collective primitives rely on successive point-to-point inter-tile transfers, leading to high communication latency.
In contrast, \glspl{noc} with hardware-supported collective communication primitives establish direct, optimized communication paths, significantly reducing communication overhead \cite{krishna2011}.
For example, consider multicasting a message of size \( \alpha \) to a chain of \( N \) clusters, demonstrated in Fig. \ref{fig_arch}.
Given an L1-to-NoC router latency of \( L_d \), router-to-router latency of \( L_r \), and a router link bandwidth of \( \beta \), the communication latency for software-based collective primitives is \( N \left(\frac{\alpha}{\beta} + 2L_d + \frac{N+1}{2}L_r \right) \).
In contrast, \glspl{noc} with hardware-supported collective communication primitives employ a path-based forwarding strategy. Each packet injected from the source node is duplicated and forwarded in-flight along the multicast-aware routing path, as illustrated in Fig. \ref{fig_arch}.
This optimization reduces the communication latency to  
\( \frac{\alpha}{\beta} + 2 L_d + NL_r \).
For example, when $\alpha=16 \: KB, \ \beta=128 \: B/cycle, \ L_d=10 \: cycles, \ L_r= 4 \: cycles, \ N= 7$, the multicast latency is reduced by $6.1\times$.

\section{FlatAttention Dataflow}
\label{sec:impl}

\begin{figure}[t!]
  \centering
  \includegraphics[width=1\linewidth, trim={2.5cm 9cm 4cm 8.25cm}, clip]{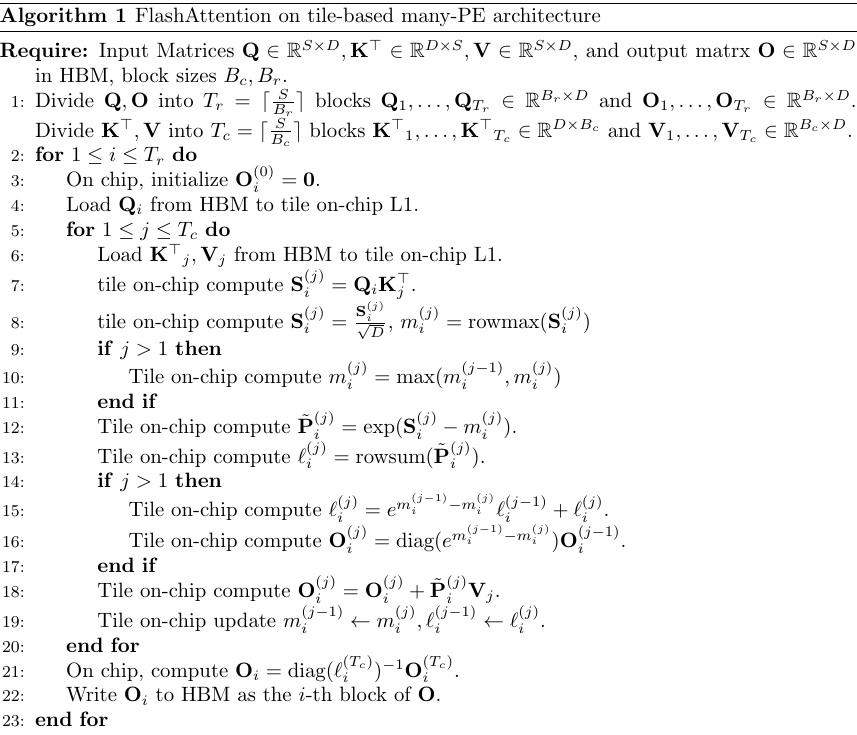}
  \vspace{-10mm}
  \label{alg:flash} 
\end{figure}

\begin{figure*}[t!]
    \centering
    \subfloat{{\includegraphics[height=5.1cm, trim={0.05cm 2.8cm 3.2cm 2.4cm}, clip]{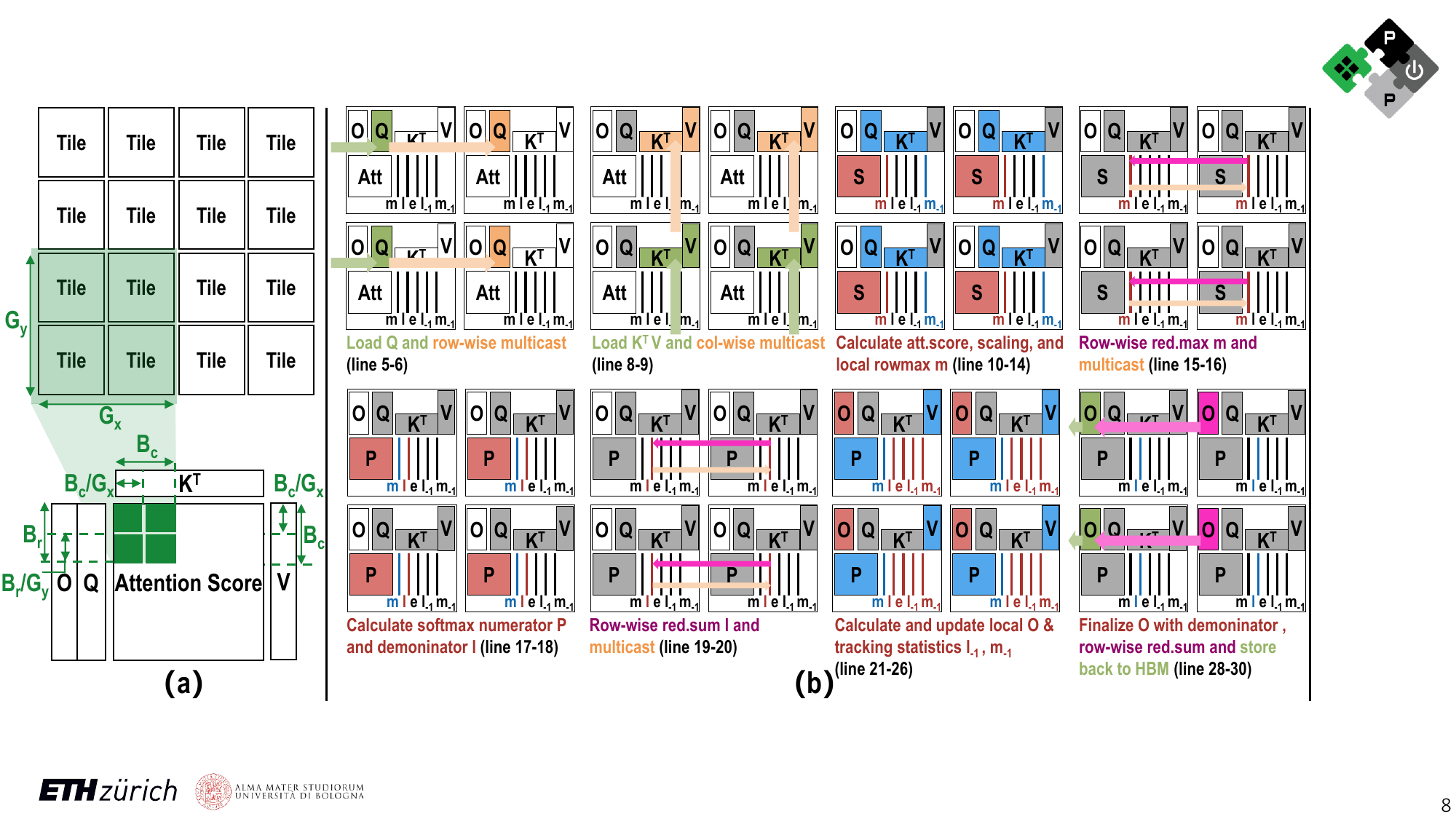}}}%
    \subfloat{{\includegraphics[height=5.1cm, trim={8cm 2.8cm 3.2cm 2.3cm}, clip]{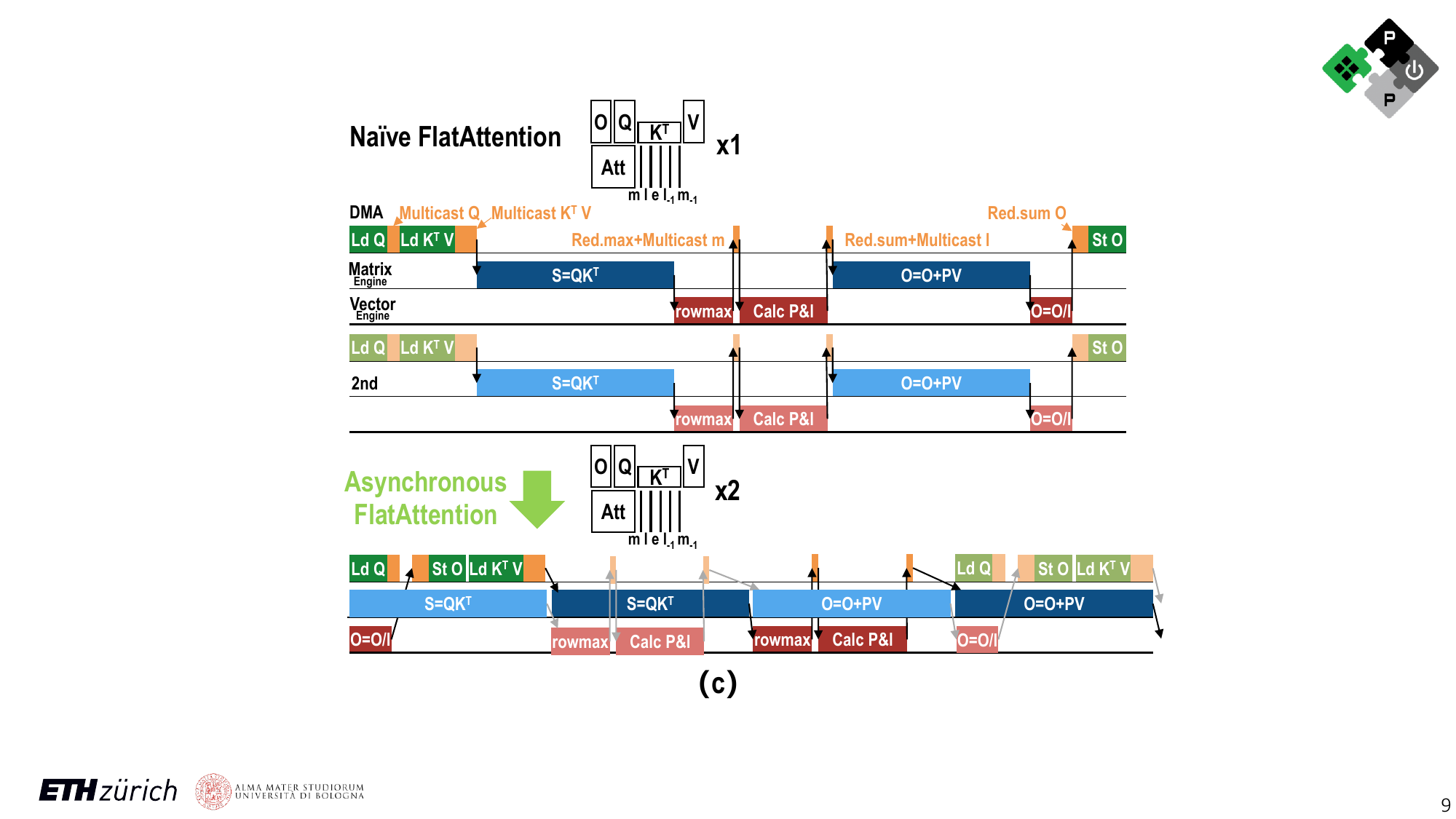}}}%
    \caption{(a) Parametric definition of FlatAttention. (b) Detailed FlatAttention dataflow, with each step corresponding to the line numbers in Algorithm 2. (c) FlatAttention dataflow optimization.}
    \label{fig_flat}
    \vspace{-6mm}
\end{figure*}

\begin{figure}[t!]
  \centering
  \includegraphics[width=1\linewidth,trim={2.5cm 6cm 4cm 5.12cm}, clip]{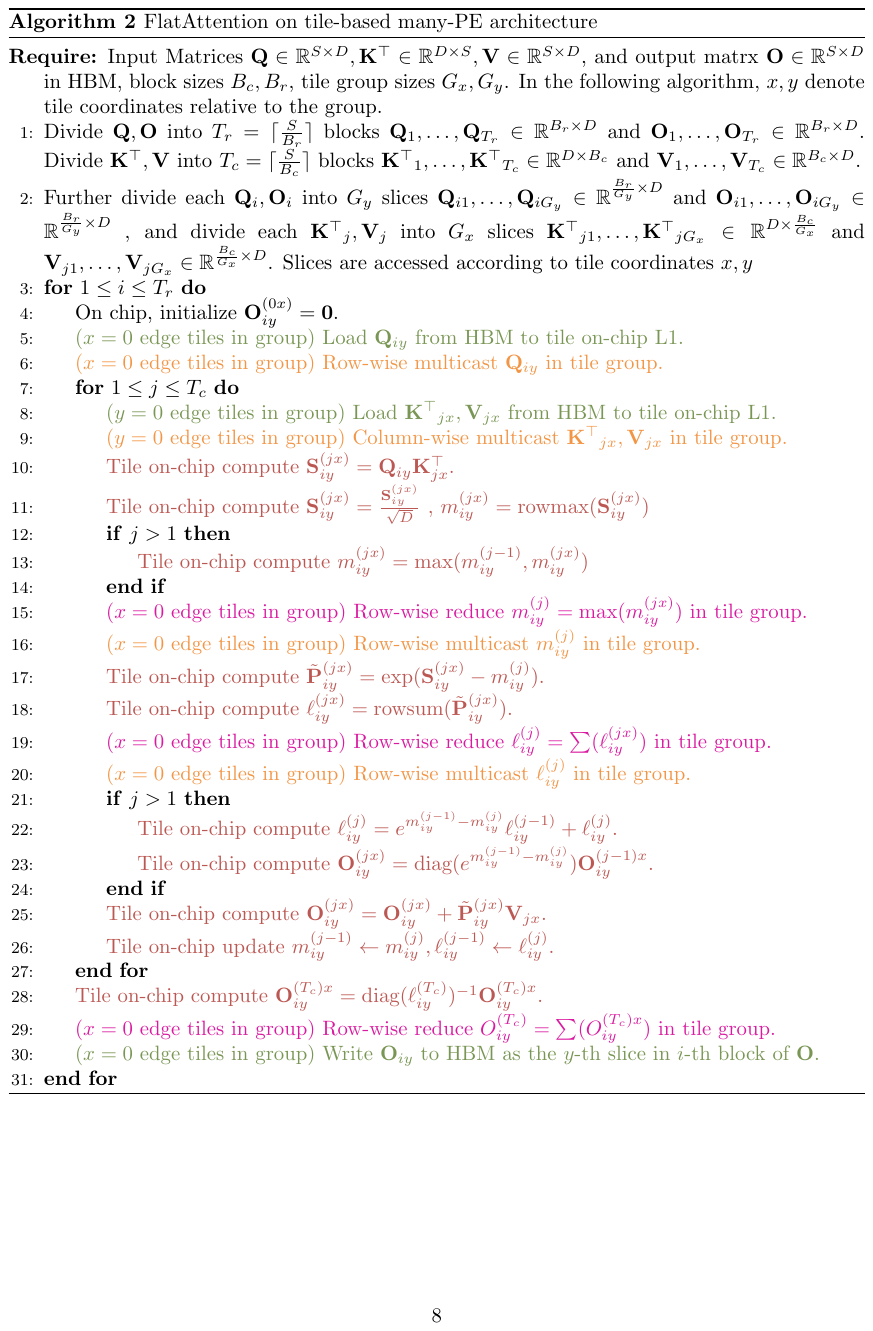}
  \vspace{-10mm}
  \label{alg:flat}
\end{figure}

\subsection{Motivation}
\label{subsec:moti}

As both FlashAttention-2 and FlashAttention-3 share the dataflow introduced in FlashAttention-2, we refer to it in the following.
Algorithm 1 outlines the FlashAttention-2 algorithm\footnote{To simplify the dataflow for tile-based many-PE accelerators, we assume the $K$ matrix is pre-transposed in HBM while still accounting for the pre-transposition time when comparing to FlashAttention on H100 for a fair comparison \cite{shah2024}.} for each head, designed to operate directly on tile-based many-PE architectures. The \gls{mha} workload is partitioned over the batch, number of heads, and output sequence length dimensions, and these blocks are distributed to the tiles where they can be processed in parallel. With this mapping, every tile processes distinct data, which it can independently access in \gls{hbm}. No communication between tiles is required, but at the same time no reuse of data across tiles is exploited.

With sequnce length $S$, head dimension $D$, number of heads $H$, batch size $B$ and block size $M \coloneqq B_r = B_c$, this dataflow results in an HBM I/O complexity of:
\vspace{-1mm}
\[
\text{IO} = 2 \cdot H \cdot B \cdot D \cdot S \cdot \left( 1 + \frac{S}{M} \right).
\]

While all other parameters are fixed by the computation, the block size parameter $M$ can be increased to lower the I/O complexity. Intuitively, larger blocks favor the reuse of data in the L1 memory of a tile.
However, the block size is constrained by the L1 memory of a \textit{single} tile, which must be able to simultaneously host tensors $Q_i, K^T_j, V_j, O_i$, at any given time.

With the goal of further reducing off-chip I/O accesses, we propose the FlatAttention dataflow, which fundamentally redefines how \gls{mha} is parallelized on tile-based architectures. FlatAttention leverages multiple tiles as a unified entity to process an \gls{mha} block, as defined above, of a significantly larger size, given that the aggregate L1 memory of a group of tiles can now be used to collectively store the block.
When $N$ tiles are grouped together, the resulting I/O complexity becomes:
\vspace{-1mm}
\[
\text{IO} = 2 \cdot H \cdot B \cdot D \cdot S \cdot \left( 1 + \frac{ S}{\textcolor{red}{\sqrt{N}} \cdot M} \right).
\]

For example, when $S = 4096$, $M = 128$, and $N = 64$, this results in a  $6.6\times$ theoretical reduction in HBM accesses.

\subsection{Detailed FlatAttention Dataflow}
\label{subsec:flat}

The FlatAttention dataflow is depicted in Fig.~\ref{fig_flat}b. We refer to a set of tiles collectively processing a block, as previously introduced, as a \textit{group}, demonstrated in Fig.~\ref{fig_flat}a. We define the shape of the group as $G_x \times G_y$. FlatAttention applies the same tiling and mapping scheme to groups as FlashAttention applies to tiles but introduces a secondary level of blocking within each group. This secondary blocking divides the $\{B_c, B_r\}$ block dimensions into smaller \textit{slices} based on the group shape $\{G_x, G_y\}$, resulting in $\{\frac{B_c}{G_x}, \frac{B_r}{G_y}\}$ slice sizes for every tile.

Algorithm 2 outlines the FlatAttention dataflow.
At a high level, the algorithm is conceptually similar to FlashAttention (Algorithm 1): different groups process distinct data, so no communication between groups is required. However, distributing the computation of an \gls{mha} block to tiles in a group introduces distinct data movement patterns within the group:

\begin{itemize}
    \item \textbf{Loading and Multicasting}: 
    Only tiles on the west edge of the group load $Q$ slices from HBM (line \tikz[baseline, yshift=4pt]{\node[shape=circle, draw, fill=color_ls, text=white, inner sep=0.5pt, font=\small] {5};}), followed by multicasting $Q$ slices row-wise \tikz[baseline, yshift=4pt]{\node[shape=circle, draw, fill=color_mc, text=white, inner sep=0.5pt, font=\small] {6};} to the other tiles in the group.
    When entering the inner loop, the tiles on the south edge load $K^T$ and $V$ slices from HBM \tikz[baseline, yshift=4pt]{\node[shape=circle, draw, fill=color_ls, text=white, inner sep=0.5pt, font=\small] {8};} followed by multicasting them column-wise \tikz[baseline, yshift=4pt]{\node[shape=circle, draw, fill=color_mc, text=white, inner sep=0.5pt, font=\small] {9};}.

    \item \textbf{Computing Attention ($Q \cdot K^T$) and Rowmax}:
    Each tile computes a segment of the attention score matrix \tikz[baseline, yshift=4pt]{\node[shape=circle, draw, fill=color_cp, text=white, inner sep=0.5pt, font=\footnotesize] {10};}. During the computation of row-wise maxima for Softmax, tiles compute partial row maxima locally\tikz[baseline, yshift=4pt]{\node[shape=circle, draw, fill=color_cp, text=white, inner sep=0.5pt, font=\footnotesize] {11};} updated with the tracking maxima\tikz[baseline, yshift=4pt]{\node[shape=circle, draw, fill=color_cp, text=white, inner sep=0.5pt, font=\footnotesize] {13};}, followed by a row-wise reduction within the group to calculate the global row maxima\tikz[baseline, yshift=4pt]{\node[shape=circle, draw, fill=color_rd, text=white, inner sep=0.5pt, font=\footnotesize] {15};}. The results are then multicasted row-wise to ensure that each tile holds the global row maxima \tikz[baseline, yshift=4pt]{\node[shape=circle, draw, fill=color_mc, text=white, inner sep=0.5pt, font=\footnotesize] {16};}.

    \item \textbf{Softmax Denominator}:
    After computing the partial Softmax denominator locally with global row maxima\tikz[baseline, yshift=4pt]{\node[shape=circle, draw, fill=color_cp, text=white, inner sep=0.5pt, font=\footnotesize] {17};}\tikz[baseline, yshift=4pt]{\node[shape=circle, draw, fill=color_cp, text=white, inner sep=0.5pt, font=\footnotesize] {18};}, the same reduction \tikz[baseline, yshift=4pt]{\node[shape=circle, draw, fill=color_rd, text=white, inner sep=0.5pt, font=\footnotesize] {19};} and multicast \tikz[baseline, yshift=4pt]{\node[shape=circle, draw, fill=color_mc, text=white, inner sep=0.5pt, font=\footnotesize] {20};} procedure applies to computing the global denominator, which is then updated with the tracking maxima and denominator \tikz[baseline, yshift=4pt]{\node[shape=circle, draw, fill=color_cp, text=white, inner sep=0.5pt, font=\footnotesize] {22};}.

    \item \textbf{Output Matrix ($O$)}:
    Each tile updates local $O$ slices and tracking statistics in the inner loop, and computes partial results for $O$ slices on exit \tikz[baseline, yshift=4pt]{\node[shape=circle, draw, fill=color_cp, text=white, inner sep=0.5pt, font=\footnotesize] {23};}-\tikz[baseline, yshift=4pt]{\node[shape=circle, draw, fill=color_cp, text=white, inner sep=0.5pt, font=\footnotesize] {28};}. FlatAttention then performs a row-wise reduction of $O$ slices \tikz[baseline, yshift=4pt]{\node[shape=circle, draw, fill=color_rd, text=white, inner sep=0.5pt, font=\footnotesize] {29};} followed by storing $O$ slices in \gls{hbm} \tikz[baseline, yshift=4pt]{\node[shape=circle, draw, fill=color_ls, text=white, inner sep=0.5pt, font=\footnotesize] {30};} only from west-edge tiles.
\end{itemize}

These communication requirements are a direct result of FlatAttention's parallelization scheme, which enables minimizing costly \textit{global} off-chip I/O by exploiting on-chip data reuse across tiles through \textit{local} on-chip communication. This trade-off of global for local requirements enables FlatAttention to achieve better scalability and performance compared to FlashAttention methods for tile-based many-PE architectures, as long as local on-chip communication is efficiently handled, as will be discussed in Section \ref{subsec:exp1}.

\subsection{Asynchronous FlatAttention}
\label{subsec:opti}

In the na\"ive version of FlatAttention (Algorithm 2), data movement and SoftMax-related computations still account for a significant portion of the runtime, as illustrated in Fig.~\ref{fig_flat}c. This reduces the overall utilization, as the system's peak performance is primarily determined by the matrix engine, which has much higher computational power compared to the vector engine. To further improve utilization, we propose leveraging the asynchronous nature of \gls{dma}, vector and matrix engine invocations to overlap the runtime of data movement and SoftMax operations with matrix multiplications.

The optimized dataflow schedules the computation of two heads concurrently on each group. While the matrix engine processes matrix multiplications for one head, the \gls{dma} and vector engine perform data movement and SoftMax operations for the other\footnote{The same optimization can be applied with two output row blocks $O_i$ instead of two heads, reducing memory requirements as the $K^T_j$ and $V_j$ blocks are shared. To simplify the evaluation, where sufficient row blocks are not available in all configurations, we adopt the presented implementation.}. Fig.~\ref{fig_flat}c demonstrates this optimization, showcasing how it can ensure the matrix engine remains nearly fully active, provided that the matrix multiplications runtime overlaps completely with data movement and SoftMax operations. Notably, FlashAttention-3 employs the same technique to improve over FlashAttention-2.

\section{Modeling and Analysis Methodology}
\label{sec:meth}

We developed a modeling and simulation framework for tile-based many-PE accelerators on the GVSoC event-based simulator \cite{gvsoc} for functional and performance simulation. GVSoC is open source and released with models for the Snitch\cite{zaruba2020snitch} single-issue RISC-V core, the Spatz\cite{perotti2025spatz} vector engine supporting the \gls{rvv} extension, the iDMA\cite{benz2023high} engine, and tile-local L1 memory and interconnect.
To extend these capabilities, we developed and calibrated new models for the RedMulE\cite{tortorella2023RedMulE} matrix engine and the FlooNoC \cite{fischer2025floonoc} fabric according to their open-source RTL implementations. The NoC model incorporates both software and hardware-supported collective communication primitives, as presented in Section \ref{sec:arch}, for design space exploration. Specifically, we can simulate hardware support for row-wise and column-wise multicast, sum-reduction, and max-reduction operations. We also extended Spatz with a custom \gls{rvv} instruction for exponential operations and a dedicated exponential unit within the FPU.
Furthermore, we integrated the DRAMSys \cite{jung2015dramsys} simulator into GVSoC for HBM modeling. i.e., we used the HBM2e specification with 64 GB/s bandwidth per channel.

Using these building blocks, we constructed the \textbf{SoftHier} model and analysis framework: a flexible, parameterizable tile-based many-PE accelerator simulator with functionality and performance models calibrated on cycle-accurate simulations of open-source RTL. SoftHier is configurable using architecture configuration files, enabling the instantiation of specific accelerator designs, e.g. to explore different numbers of \glspl{ce} in the RedMulE units.

In the SoftHier framework, we implemented the FlashAttention and FlatAttention dataflows in C, incorporating APIs for matrix engine offload and DMA engine inter-tile communication, along with \gls{noc} collective primitives. The software was compiled using the \texttt{GNU RISC-V GCC} compiler with \texttt{-O3} optimization. We assume a 1 GHz clock frequency, and preheat every tile's instruction cache at the start of the simulation. For FlashAttention, we parallelize across the batch, number of heads and output sequence length dimensions to ensure that all tiles are utilized. For both implementations, we select the slice size per tile to maximize local L1 memory occupancy while maintaining a square configuration, i.e., $\frac{B_r}{G_y} = \frac{B_c}{G_x}$.

\section{Experimental Results}
\label{sec:expe}

\subsection{FlashAttention vs. FlatAttention}
\label{subsec:exp1}

\begin{table}[ht]
    \centering
    \caption{System Specifications}
    \label{tab:system_specs}
    \renewcommand{\arraystretch}{1.2} 
    \setlength{\tabcolsep}{8pt} 
    \begin{tabular}{@{}l@{ }l@{}}
        \toprule
        \textbf{System} & 32×32 Tiles, 1024-bit NoC link width \\ 
        \midrule
        \textbf{HBM} & 16x2 Channels, equally divided over west and south edges \\ 
        \textbf{Tile} & \textbf{RedMulE Matrix Engine}: 32×16\,CE\,array, 1\,TFLOPS@FP16 \\ 
        & \textbf{Spatz Vector Engine}: 16\,FPU, 128\,GFLOPS@FP16 \\ 
        & \textbf{Local Memory}: 384 KB, 512 GB/s \\ 
        \textbf{Summary} & 1024\,TFLOPS Peak\,Performance\,, 2\,TB/s \,Peak\,HBM\,Bandwidth \\ 
        \bottomrule
    \end{tabular}
    \vspace{-3mm}
\end{table}

\begin{figure}[t!]
  \centering
  \includegraphics[width=\columnwidth]{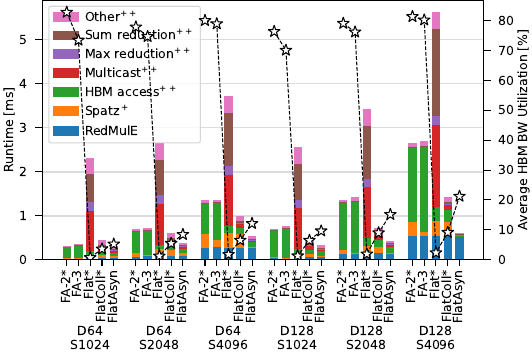}
  \caption{Runtime breakdown (bars) and average HBM BW utilization (star markers) for different \gls{mha} implementations and layer sizes. $^+$Runtime not overlapped with RedMulE. $^{++}$Runtime not overlapped with either Spatz or RedMulE. *Implementations without double buffering.}
  \label{plot1}
  \vspace{-6mm}
\end{figure}

We first compare different MHA implementations and layer sizes on a given \arch accelerator configuration, as specified in Table~\ref{tab:system_specs}.
We evaluate both FlashAttention-2 (\textit{FA-2}) and FlashAttention-3 (\textit{FA-3}) implementations on \arch accelerators, where \textit{FA-3} introduces a similar scheduling mechanism as presented in Section \ref{subsec:opti}.
For FlatAttention, we set the group size to include all tiles in the system, i.e. $G_x=G_y=32$.
We evaluate a na\"ive implementation without (\textit{Flat}) and with (\textit{FlatColl}) hardware support for efficient collective primitives on the \gls{noc}, as well as the optimized FlatAttention dataflow (\textit{FlatAsyn}) described in Section \ref{subsec:opti} with \gls{noc} collective primitives. We evaluate multiple MHA layers, varying the sequence length $S \in \{1024, 2048, 4096\}$ and head dimension $D \in \{64, 128\}$, while fixing batch size $B = 2$ and heads $H = 32$.

Fig.~\ref{plot1} presents the runtime breakdown and average HBM bandwidth utilization. We observe that FlashAttention exhibits a highly memory-bound behavior on the target \arch system, with HBM bandwidth utilizations reaching up to \FlashBW\% on average. HBM access is the dominant runtime component, limiting overall compute utilization. Even with \textit{FA-3}’s optimized dataflow for overlapping matrix multiplication and Softmax operations, the saturated HBM bandwidth prevents further speedup. Additionally, \textit{FA-3} introduces an overhead for more complex scheduling.

\textit{Flat} significantly reduces HBM access time compared to \textit{FA-3} due to the decreased I/O complexity.
However, software-based collective primitives used in \textit{Flat} rely on successive point-to-point inter-tile transfers. For instance, a row-wise multicast in this configuration requires 31 sequential unicast transmissions, incurring substantial on-chip communication overhead and resulting in worse performance than FlashAttention.
Instead, with efficient collective primitives enabled on the \gls{noc} (\textit{FlatColl}), on-chip inter-tile communication is significantly accelerated, leading to better performance than FlashAttention across most MHA layers. The \textit{FlatAsyn} implementation shows that we can further improve performance by overlapping SoftMax, data movement, and matrix multiplication operations.
Overall, our optimizations result in up to \OptFlatPerfSU× speedup and \OptFlatHBMTraficSave× HBM traffic reduction over \textit{FA-3} (D128, S4096).

\subsection{Tile Group Scale Trade-offs for FlatAttention}
\label{subsec:exp2}

\begin{figure}[t!]
  \centering
  \includegraphics[width=\columnwidth]{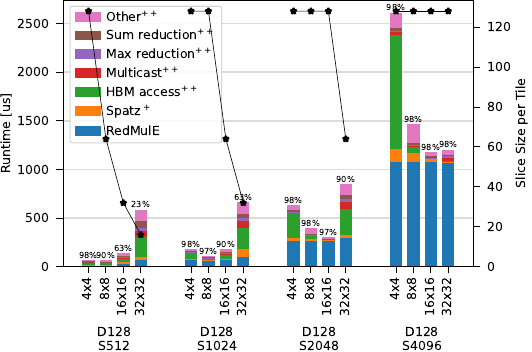}
  \caption{Runtime breakdown for different (square) flattening scales and layer sizes. Percentage labels above the bars indicate the average utilization of the RedMulE units when active. $^+$Runtime not overlapped with RedMulE. $^{++}$Runtime not overlapped with either Spatz or RedMulE.
  }
  \label{plot2}
  \vspace{-6mm}
\end{figure}

Although \textit{FlatAsyn} achieves the best performance across all MHA layers in Fig.~\ref{plot1}, it does not fully optimize utilization for shorter sequence lengths such as 2048 and 1024, where RedMulE runtime cannot completely overlap with other operations. To determine the optimal performance configuration for FlatAttention, we analyze the impact of different (square) group sizes $G_x, G_y \in \{4, 8, 16, 32\}$, on the specific \arch accelerator configuration defined in Table~\ref{tab:system_specs}.

We evaluate multiple MHA layers, varying the sequence length $S \in \{512, 1024, 2048, 4096\}$, while fixing $D=128$, $H=32$ and $B=4$.

Fig.~\ref{plot2} presents the runtime breakdown and \gls{mha} workload slice size per tile, i.e. $\frac{B_r}{G_y} = \frac{B_c}{G_x}$, across different group scales. For long sequence lengths such as 4096, the slice size per tile remains constant due to L1 memory capacity. In this case, increasing the group size reduces overall HBM I/O complexity, as demonstrated in Section~\ref{subsec:flat}, leading to reduced HBM access runtimes and improved overlap with matrix multiplication on RedMulE. The 16×16 and 32×32 group scales achieve 88\% and 87\% utilization, respectively, for a sequence length of 4096.

However, for shorter sequence lengths such as 512, the situation is different. As the group scale increases, the slice size per tile decreases due to the fixed sequence length, introducing two performance overheads: 

\begin{itemize}
    \item \textbf{Reduced RedMulE utilization}: 
    Smaller slices per tile lead to lower RedMulE utilizations. For example, in a 32×32 group with a sequence length of 512, every tile’s RedMulE achieves only 23\% utilization when active.

    \item \textbf{Increased synchronization overhead}:
    Smaller slices per tile result in shorter RedMulE runtimes. As a result, constant overheads associated with synchronization and data movement, such as \gls{hbm} access latency ($\sim$200 cycles), constitute a larger fraction of the overall runtime.
\end{itemize}

We refer to this effect as \textit{over-flattening}. For moderate sequence lengths, both effects occur simultaneously: larger group scales effectively reduce I/O complexity but also introduce the risk of over-flattening. For every sequence length, there exists an optimal group scale balancing the two effects.

\subsection{Co-exploration of Architecture and Algorithm Parameters}
\label{subsec:exp3}

Lastly, we demonstrate how the SoftHier framework can be used to identify an optimal \arch architecture configuration. Our goal is to design a tile-based accelerator with comparable peak performance to Nvidia's H100
(989 TFLOPS FP16/BF16 without sparsity)
while improving utilization and reducing overall HBM bandwidth requirements on \gls{mha} workloads. We evaluated a set of candidate architecture configurations with varying NoC fabric granularity and HBM channel connectivity. Table~\ref{tab:fabric_specs} presents the tile specifications as a function of fabric granularity, ensuring constant peak system performance (1024 TFLOPS) and on-chip memory capacity. For every accelerator candidate, we evaluate multiple \gls{mha} layers, searching for optimal performance across different dataflow implementations, including FlashAttention-3 and FlatAttention with varying square-shaped group sizes.

\begin{figure}[t]
    \centering
    
    \subfloat{
        \includegraphics[width=0.45\columnwidth]{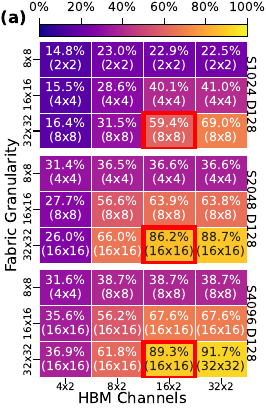}%
        \label{fig:left}
    }
    \begin{minipage}[b]{0.45\columnwidth}
        \centering
        
        \subfloat{
            \includegraphics[width=\columnwidth, height=0.77\columnwidth]{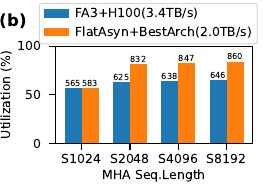}%
            \label{fig:top_right}
       }\\[2pt] 
        
        \subfloat{
            \includegraphics[width=\columnwidth, height=0.77\columnwidth]{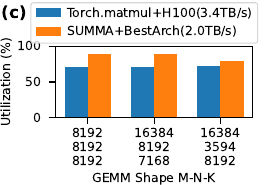}%
            \label{fig:bottom_right}
        }
        
    \end{minipage}
    
    \caption{(a) Heatmap of utilization with best group size. (b) Comparison with FlashAttention-3 on H100 solutions, with absolute performance in TFLOPS labeled above the bar. (c) Comparison of GEMMs, including FFN layer in Meta's LLaMA 70B \cite{gemmbench}, between \textit{BestArch} and H100.}
    \label{plot3}
    \vspace{-6mm}
\end{figure}

Based on the results shown in Fig.~\ref{plot3}a, we can select a configuration (\textit{BestArch}) that optimizes for performance over cost, featuring a 32×32 fabric granularity and 16×2 HBM channels. We then compare its performance directly against FlashAttention-3, based on the H100 performance numbers in Shah et al.\cite{shah2024}, using the same \gls{mha} layers while also accounting for the $K$ matrix pre-transposition time in FlatAttention for fair comparison.
In Fig.~\ref{plot3}b, the \textit{BestArch} configuration with FlatAttention achieves up to \SUtoNV× higher utilization while requiring \BWreducetoNV\% less HBM bandwidth compared to the H100 GPU.
Beyond \gls{mha}, common GEMM kernels utilizing the collective-based SUMMA dataflow \cite{van1997summa
} on \textit{BestArch} also achieve up to \GEMMSUtoNV× higher utilization over H100 \cite{gemmbench} in Fig.~\ref{plot3}c.
Using the \gls{ge} reported for the individual components \cite{zaruba2020snitch, perotti2025spatz, benz2023high, tortorella2023RedMulE, fischer2025floonoc}, we estimated the die size of \textit{BestArch} in TSMC 5nm technology, the same used by the H100.
Considering 4 transistors per \gls{ge}, a transistor density of 138.2\,MTr/mm\textsuperscript{2}, an SRAM bit-cell size of 0.021\,\textmu m\textsuperscript{2}, and assuming 66\% area utilization, \textit{BestArch} features a die size of 457\,mm\textsuperscript{2}, enabling a \DieSave× reduction to H100 GPU.

\begin{table}[!t]
    \centering
    \caption{Fabric Granularity and Tile Specifications}
    \label{tab:fabric_specs}
    \renewcommand{\arraystretch}{1.2} 
    \setlength{\tabcolsep}{8pt} 
    \begin{tabular}{lccc}
        \toprule
        \textbf{Fabric Granularity} & \textbf{32×32} & \textbf{16×16} & \textbf{8×8} \\ 
        \midrule
        \textbf{RedMulE CE Array} & 32×16 & 64×32 & 128×64 \\ 
        \textbf{Spatz FU Count} & 16 & 64 & 256 \\ 
        \textbf{Local Memory Size (KB)} & 386 & 1526 & 6144 \\ 
        \textbf{Local Memory Bandwidth (GB/s)} & 512 & 2048 & 8192 \\ 
        \bottomrule
    \end{tabular}
    \vspace{-4mm}
\end{table}

\section{Conclusion}

\vspace{-2mm}
We propose FlatAttention, an optimized dataflow for \gls{mha} on \arch accelerators, co-designed with \gls{noc} collective primitives, achieving up to \MaxUti\% utilization while reducing HBM traffic by \OptFlatHBMTraficSave× compared to FlashAttention-3 dataflow, on tile-based accelerators. 
Through algorithm-architecture co-exploration, an optimal accelerator configuration matching the peak performance of Nvidia's H100 GPU is selected, which requires \BWreducetoNV\% less HBM bandwidth than H100 and \DieSave× reduction in die size, with FlatAttention achieving up to \SUtoNV× higher utilization compared to FlashAttention-3 on H100,  and its GEMM reaching up to \GEMMSUtoNV× higher utilization over H100.
Future work includes end-to-end LLM inference on multi-chiplet systems with 3D-stacked memory.

\vspace{-2mm}
\section*{Acknowledgment}
This work is supported by the ETH Future Computing Laboratory (EFCL) and Huawei ZRC.

\bibliography{paper}
\bibliographystyle{IEEEtran}

\end{document}

%% file: glossary.tex
\newacronym{hbm}{HBM}{High Bandwidth Memory}
\newacronym{llm}{LLM}{Large Language Model}
\newacronym{tdp}{TDP}{Thermal Design Power}
\newacronym{dnn}{DNN}{Deep Neural Network}
\newacronym{mha}{MHA}{Multi-Head Attention}
\newacronym{ml}{ML}{Machine Learning}
\newacronym{pe}{PE}{Processing Element}
\newacronym{ce}{CE}{Compute Element}
\newacronym{soa}{SoA}{state-of-the-art}
\newacronym{dma}{DMA}{Direct Memory Access}
\newacronym{noc}{NoC}{Network on Chip}
\newacronym{fu}{FU}{Function Unit}
\newacronym{ai}{AI}{artificial intelligence}
\newacronym{rvv}{RVV}{RISC-V Vector}
\newacronym{ge}{GE}{Gate Equivalent}
\newacronym{FA}{FA}{FlashAttention}
\newacronym{gemm}{GEMM}{General Matrix Multiplication}